\documentclass[final,5p,times,twocolumn]{elsarticle}

\usepackage{lineno}
\usepackage{graphicx}

\usepackage[integrals]{wasysym} 

\modulolinenumbers[5]

\journal{Spectrochimica Acta Part B: Atomic Spectroscopy}

\usepackage{amssymb, hyperref}
\usepackage{amsmath}
\usepackage{lineno}

\begin{document}

\begin{frontmatter}

\title{Spectroscopy of resonantly saturated selective reflection from high-density rubidium vapor using pump-probe technique}

\author[JIHT]{Vladimir Sautenkov}
\author[JIHT,HSE]{Sergey Saakyan\corref{cor}}%
\ead{saakyan@ihed.ras.ru}
\author[JIHT]{Andrei Bobrov}
\author[JIHT]{Leonid Khalutornykh}
\author[JIHT]{Boris B. Zelener}

\address[JIHT]{Joint Institute for High Temperatures, Russian Academy of Sciences (JIHT RAS), Izhorskaya St. 13 Bld. 2, Moscow 125412, Russia}
\address[HSE]{National Research University Higher School of Economics (NRU HSE), Myasnitskaya Ulitsa 20, Moscow 101000, Russia}
\cortext[cor]{Corresponding author}

\begin{abstract}
    We study the resonant saturation of selective reflection from the interface between a transparent dielectric and high-density rubidium vapor on the D$_2$-line. According to our estimates in the selected atomic density range, the dipole–dipole self-broadening of the line can vary from $13.2$ to $39.6$~GHz. Two tunable lasers are used as sources of pump and probe beams with orthogonal linear polarizations. The selective reflection spectra of the probe laser beam are studied at different atomic densities and at pump beam intensities from $0$ to $8.8$~kW$\cdot$cm$^{-2}$. At high pump intensities, narrow structures are observed around the pump beam frequency. The narrow structures are associated with power broadening effects. Increase of the pump intensity reduces the spectral width and the magnitude of the selective reflection resonances. The intensity dependence of the width and the magnitude are measured. By changing the pump intensity, it is possible to control the spectral width and reflectivity. 
\end{abstract}


\end{frontmatter}


\section{Introduction}
High-density atomic vapors, where dipole–dipole interactions induce self-broadening of resonance transitions~\cite{lewis1980collisional} dominated, can be a resource of new nonlinear optical effects. For example, it is shown in~\cite{friedberg1973frequency} that in an excited atomic gas, the value of the nonlinear Lorentz local shift depends upon the degree of population inversion between the two interacting levels $\eta$ (the fractional population difference between the ground state and the excited state). The classic Lorentz shift in the high-density potassium vapor~\cite{maki1991linear} and in the high-density rubidium vapor~\cite{wang1997selective} is measured by using selective reflection (SR) spectroscopic techniques. The first observation of the predicted nonlinear Lorentz shift and an unexpected reduction of the dipole induced broadening of the resonance transitions in an incoherently excited atomic rubidium vapor was reported in~\cite{sautenkov1996dipole}. In the experiment, the probe laser and the pump laser were applied. The frequency of the probe laser was scanned over the resonance transition 5S$_{1/2}$–-5P$_{3/2}$ in rubidium atoms and SR spectra were recorded. A frequency of the pump laser was detuned to the far wing of the transition, where an absorption length was much great than the wavelength ($l_{\text{abs}}$$\gg$$\lambda$). Optical off-resonant excitation of rubidium atoms was incoherent due to the radiation trapping effect~\cite{holstein1951imprisonment,sautenkov2020spectral}. On the basis of comparison of experimental and calculated curves it was suggested that the dipole self-broadening is linearly dependent on the ground-state population of atoms. The observed excitation dependence of the dipole self-broadening was explained by the quasi-static mechanism of the dipole–dipole interaction in the high-density atomic vapor~\cite{leegwater1994self}. We would like to note that there is a linear relation between the fractional population difference $\eta$ and the ground state population $N_\text{g}$~\cite{friedberg1973frequency,van1999measurement}. The linear dependence of the Lorentz local shift and self-broadening on the fractional population difference $\eta$ were measured in an incoherently excited potassium vapor~\cite{van1999measurement}. The linear relation between the self-broadening and the fractional population difference $\eta$ was confirmed also by experiments involving rubidium~\cite{li2009excitation} and cesium vapors~\cite{meng2016excitation}.

Remarkable differences exist between the incoherent excitation, where laser frequency is detuned to the far wing of a spectral line, and the resonant excitation, where laser frequency is tuned to the center of a spectral line. In case of the resonance excitation, the radiation trapping effect can be neglected due to the short absorption length ($l_{abs}$$<$$\lambda$). Under this condition, the effective lifetime of the excited state population is less than the radiative decay time of the optical excitation. 

In the resonance pump-probe experiment involving cesium vapor, narrow resonances in dipole broadened reflection spectra are observed~\cite{sautenkov1997observation}. The resonances appear due to the four-wave mixing of the laser beams at a self-induced population grating. The measured spectral widths allowed to estimate the relaxation time of population of the atomic excited state. The signal was recorded with the similar polarized laser beams. Due to lack of the interference of the beams, the signal did not appear in case of orthogonally polarized pump and probe laser beams. Recently the coherent resonances were observed in a pump-probe experiment involving a high density rubidium vapor~\cite{sautenkov2023coherent}. 

Also, a study of resonance excitation of the high-density potassium vapor was performed by using two lasers in a V-level scheme~\cite{sautenkov2008observation}. The probe laser frequency was scanned over the 4S$_{1/2}$--4P$_{1/2}$ transition. The pump laser frequency was tuned inside the central part of the transition 4S$_{1/2}$--4P$_{3/2}$. In the selected V-scheme chosen, the Autler–Towns doublets appeared due to the coherent mode of a strong pump excitation.

Recently we studied the resonance optical saturation of SR from the high-density rubidium atomic vapor by using one tunable laser only~\cite{sautenkov2021optical,bobrov2021dipole,sautenkov2022spectral}. Reduction of the reflection coefficient and the spectral width of the resonances were observed. Stabilization of the reflection coefficient and the spectral width was observed at high pump intensities. In the presented work, experimental research of the nonlinear SR from high density rubidium vapor is conducted with orthogonally polarized pump and probe laser beams. The pump beam induces reduction of the spectral width, and reflectivity along with power broadening effects are observed and discussed.

\section{Experimental setup and methods}

Saturated SR from the window–vapor interface in the high-temperature vapor cell with natural abundance of rubidium-85 and rubidium-87 isotopes is investigated at the transition 5S$_{1/2}$–5P$_{3/2}$ (D$_2$-line). The sapphire cell with YAG windows is the same as that in~\cite{sautenkov2021optical,bobrov2021dipole,sautenkov2022spectral}. The atomic number density $N$ is determined by the temperature of the cell coldest spot.
We conduct experiments at various temperatures of the vapor cell. The variation of the temperature from 367 to 427$^\circ$C means the density range of $N$ = (1.2 -- 3.6) $\times\,10^{17}$~cm$^{-3}$~\cite{alcock1984vapour}. By using expression for dipole self-broadening width (FWHM)
\begin{equation}
    \Gamma=KN  
\end{equation}
with factor $K/2\pi=(1.10\pm0.17)\times10^{-16}$~GHz$\cdot$cm$^3$ for the working transition~\cite{kondo2006shift,weller2011absolute}, we estimate the frequency range for $\Gamma/2\pi$ as $(13.2-40)$~GHz.

The experimental configuration for our pump--probe spectroscopy employs two external-cavity diode lasers: a high-power pump laser and a probe laser. The pump laser (Toptica TA pro) is equipped with a tapered amplifier and provides more than $1.5$~W of output power. Frequency of the pump beam is measured by a wavemeter (High-Finesse/Angstrom WS-U) with an absolute accuracy of 2~MHz. Examples of applications for frequency measurements are described in~\cite{saakyan2015frequency}.

\begin{figure}[ht]
    \includegraphics[width=\linewidth]{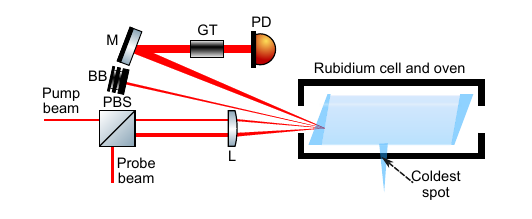}
    \caption{Experimental setup for selective reflection measurements. Two orthogonally polarized laser beams from pump and probe lasers are combined by the polarizing beam splitter (PBS) and are focused by a lens (L) with focal length of 200~mm at the internal surface of the cell window. The reflected probe beam is directed by means of a mirror (M) to the photodiode (PD) through the Glan--Thompson polarizer (GT). The reflected pump beam is dumped with the beam-block (BB).}
    \label{fig1_setup}
\end{figure}

A half-wave plate in a motorized rotation mount and a polarizing beam splitter are used for varying the intensity of the pump beam. The power of the pump beam is measured using a calibrated power meter. Frequency of the probe laser is scanned across the transition 5S$_{1/2}$--5P$_{3/2}$, achieving a mode-hop-free tuning range of 45~GHz. For absolute frequency calibration, we record the saturated absorption spectrum of a rubidium reference cell with a natural abundance of rubidium isotopes. Additionally, in order to control the mode-hop-free tuning range, we employ a reference Fabry--Perot cavity with a free spectral range of 1.2~GHz. Fig.~\ref{fig1_setup} illustrates our experimental setup designed for pump and probe SR measurements.

The pump and probe beams, orthogonally polarized, are combined by a polarizing beam splitter cube PBS, and then are focused onto the YAG--vapor interface in the vapor cell. This orthogonal polarization scheme was deliberately selected in order to prevent possible formation of narrow resonances due to the coherent scattering of probe and pump optical fields by induced oscillations of atomic populations at the beat frequency~\cite{li2009excitation,meng2016excitation}. The beam intensities are determined as $I=2P(\pi w_x w_y)^{-1}$, where $P$ is the beam power and $w_{x,y}$ denote the beam radii at the $e^{-2}$ level. The lens with a focal length of $F=200$~mm focuses the pump and probe beams into elliptical spots having dimensions of $w_{x,y}^{\text{probe}}=88$, and $56$~$\mu$m, and $w_{x,y}^{\text{pump}}=135$ and  57~$\mu$m. The probe beam intensity $I_{\text{probe}}$ is kept below 4~W$\cdot$cm$^{-2}$ in order to avoid saturation and distortion of the recorded profiles which can be induced by the probe beam. The probe beam is incident at the 64~mrad, with the 60 mrad separation angle between the pump and probe beams. The angle of incidence for the pump beam is maintained below 4~mrad. The reflected probe beam is directed through a narrow-band dielectric mirror and is then further filtered by a Glan--Thompson polarizer before reaching the photodiode. The polarizer enhances the isolation of the high-power pump beam from the photodiode. The SR signal is simultaneously recorded by a digital oscilloscope along with reference signals of the saturation absorption from a room-temperature vapor cell and a reference Fabry–Perot cavity~\cite{saakyan2015frequency}. By using a Python script, we have made the experimental setup and measurement sequences automated. This approach it possible for us to gather a comprehensive dataset with multiple SR curves. All measurements are conducted at four different values of atomic number density.

The measured spectral dependence of the SR coefficient $R$ is normalized by the non-resonant reflection from the YAG--vacuum interface $R_0\approx8.5$\% as $\delta R=(R-R_0)/R_0$. This measured value is compared with the calculated value derived from the refractive index of the garnet $n=1.821$ at $800$~nm~\cite{zelmon1998refractive}.

\section{Experimental results and discussions}

\begin{figure}[t]
    \includegraphics[width=\linewidth]{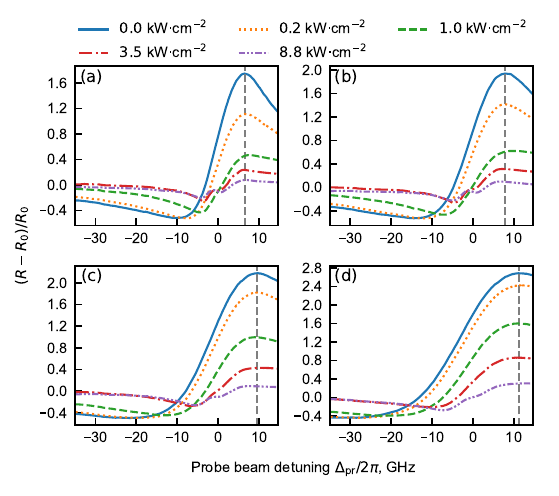}
    \caption{Spectral dependence of SR coefficient for the probe beam measured at different number densities: (a)~$N=1.2\times10^{17}$~cm$^{-3}$ ($\Gamma/2\pi=13.2$~GHz), (b)~$N=1.7\times10^{17}$~cm$^{-3}$ ($\Gamma/2\pi=18.7$~GHz), (c)~$N=2.5\times10^{17}$~cm$^{-3}$ ($\Gamma/2\pi=27.5$~GHz), and (d)~$N=3.6\times10^{17}$~cm$^{-3}$ ($\Gamma/2\pi=39.6$~GHz). Accuracy of the estimation of self-broadening $\Gamma/2\pi$ is limited by 15\% due to uncertainty of the $K$ factor.}
    \label{fig2_SR}
\end{figure}

Selected SR curves, recorded for four different atomic densities and five pump beam intensities, are shown in Fig.~\ref{fig2_SR}. The frequency scale is expressed as the probe frequency detuning $\Delta_{\text{pr}}/2\pi$ from the frequency of the rubidium-85 hf component 5S$_{1/2}(\rm{F}=3)$--5P$_{3/2}(\rm{F'}=4$)~\cite{saakyan2015frequency}. The pump laser frequency is tuned to this hf component (detuning $\Delta_{\text{pump}}/2\pi=0$).

The frequency scale is expressed as the probe frequency detuning $\Delta_{\text{pr}}/2\pi$ from the frequency of the rubidium-85 hf component 5S$_{1/2}(\text{F}=3)$--5P$_{3/2}(\rm{F'}=4)$~\cite{saakyan2015frequency}. The pump laser frequency is tuned to this hf component (detuning $\Delta_{\text{pump}}/2\pi$ = 0). The linear SR coefficient $\delta R$ is recorded with the blocked pump beam (pump intensity $I_\text{pump}$ = 0). In Fig.~\ref{fig2_SR}, the probe beam detuning corresponding to the maximum selective reflectivity of the probe beam with the blocked pump beam is indicated by dashed line. The curves, recorded with the unblocked pump beam, demonstrate strong dependence of SR spectral profiles on the pump beam intensity.

The shapes of a linear SR resonance on the D$_2$-line can be described by dispersion functions with 15\% accuracy~\cite{van1997observation}. The width of the SR resonance $\Delta\omega_{\text{SR}}$ for these resonances can be estimated as the frequency interval between the minimum and maximum of the curve~\cite{maki1991linear,van1997observation}. We used this approach for the evaluation of the spectral width of nonlinear SR profile as well~\cite{sautenkov2021optical,bobrov2021dipole}. The spectral width $\Delta\omega_{\text{SR}}$ includes intensity-dependent dipole self-broadening $\Gamma(I_\text{pump})$ and intensity-independent hf splitting of atomic transitions in Rb isotopes.

\begin{figure}[ht]
    \includegraphics[width=\linewidth]{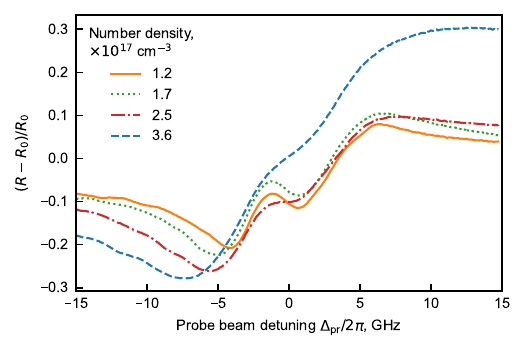}
    \caption{Spectral dependence of selective reflection coefficient for the probe beam recorded at the maximal pump intensity (Rabi frequency $\Omega_{\text{max}}/2\pi=7.5$~GHz).}
    \label{fig3}
\end{figure}

In previous works, we have shown that intensity-dependent dipole self-broadening is proportional to the ground state population $N_\text{g}$
\begin{equation}
    \Gamma(I)=KN_\text{g}
    \label{zwei}
\end{equation}	 
The spectral width of the curves decreases with the growing pump intensity. At high intensities, narrow spectral features appear near zero detuning ($\Delta_\text{pump}/2\pi$ = 0). We think that this effect can be attributed to the optical field induced coherent effects. It is more convenient to consider the optical-field-induced effects by means of the Rabi frequency $\Omega=Ed/\hbar$. In this relation, $E$ is optical electric field and $d$ is the projection of the atomic dipole moment along the field.	In our case, the Rabi frequency can be expressed by the laser beam intensity $I$ (in W$\cdot$cm$^{-2}$ units)~\cite{sautenkov2008observation,sautenkov2021optical} as $\Omega/2\pi\approx8\times10^7I^{1/2}$~Hz.		
In our experiment, the maximal pump intensity 8.8 kW$\cdot$cm$^{-2}$ corresponds to the Rabi frequency $\Omega_{max}/2\pi\approx$ 7.5~GHz. The spectral profiles of the SR coefficient near zero detuning are presented in Fig.~\ref{fig3}.

The narrow structures for number densities $1.2\times10^{17}$ cm$^{-3}$ and $2.5\times10^{17}$ cm$^{-3}$ are similar. For high densities $3.6\times10^{17}$~cm$^{-3}$ and $2.5\times10^{17}$~cm$^{-3}$, the features are smoothed. 
The recorded spectra can be explained by power broadening effects. In the textbook~\cite{laser2008laser}, a calculated profile of a homogeneously broadened transition (linewidth $\Gamma_\text{S}$) pumped by a strong pump beam fixed at a central frequency $\omega_0$ and probed by a weak tunable probe beam is discussed. The spectral profile changes for different values of the Rabi frequency $\Omega$. The pump beam induces the population modulation with the Rabi flopping frequency, and sidebands are generated. The superposition of these resonances gives the absorption spectrum depending on the ratio $\Omega/ \Gamma_\text{S}$. For a strong pump beam ($\Omega>\Gamma_\text{S}$), the frequency interval between sidebands becomes larger than their width, and a dip of the width appears at the resonance frequency $\omega_0$.

\begin{figure}[ht]
    \includegraphics[width=\linewidth]{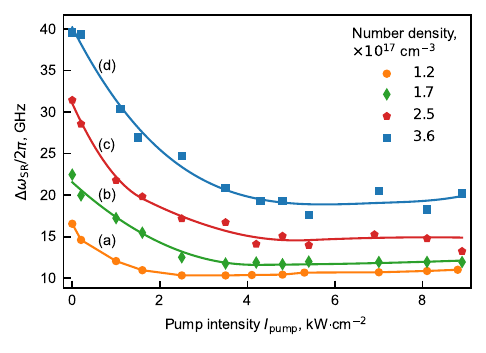}
    \caption{Width $\Delta\omega_{SR}$ of the selective reflection resonance vs. pump beam intensity $I_{\text{pump}}$ at zero pump detuning. The solid curves are guides to the eyes. Uncertainties is approximately 5\% for all points. (a) The set of measurements (orange circles and the curve) is performed at the density $1.2\times10^{17}$~cm$^{-3}$; (b) the green diamonds and the curve $1.7\times10^{17}$~cm$^{-3}$; (c) the red pentagons and the curve for $2.5\times10^{17}$~cm$^{-3}$; (d) the blue squares and the curve for $3.6\times10^{17}$cm$^{-3}$.}
    \label{fig4_width}
\end{figure}

In the experiment, the power broadening would induce a visible contribution to the width of the SR resonance $\Delta\omega_{\text{SR}}/2\pi$. Results of our measurements of the intensity-dependent width $\Delta\omega_\text{SR}/2\pi$ for four number densities are shown in Fig.~\ref{fig4_width}.       

The frequency difference between the measured $\Delta\omega_{\text{SR}}/2\pi$ with the blocked pump beam and the estimated dipole-dipole interaction induced width $KN/2\pi$ can be attributed to the non-resolved hf splitting of the resonance transitions. The results presented for the width dependence $\Delta\omega_{\text{SR}}/2\pi$ on the pump beam intensity are combinations of data and guides for eyes. The curves merely connect the data points. Obvious similarities between the curves exist. 

Fig.~\ref{fig4_width} shows that in the region of low pump intensities, it is possible to see a steep reduction of the width, and at high intensities, the data lay on a flat function quite well. The nonlinear dipole broadening $\Gamma(I)$ depends on the ground state population $N_\text{g}$ (see Eq. (\ref{zwei})). Considering this relation along with contributions of the hf structure and the power broadening of the transition allows to estimate the ground state population $N_\text{g}$. At the number density $3.6\times10^{17}$~cm$^{-3}$, the contribution of the hf splitting is small compared to the dipole broadening. We can evaluate the population $N_\text{g}$ for the flat part of the spectral width as $N/2$. The fractional difference between the ground and excited state $\eta$~\cite{friedberg1973frequency} can be estimated as well.

One of the goals of the present paper is to consider the possible optical control of the reflected probe beam by the pump beam. We measure the normalized reflection coefficient $\delta R_\text{fix}=(R - R_0)/R_0$ at the probe laser frequency detuning where the linear maximal coefficients are recorded (dashed lines in Fig.~\ref{fig2_SR}). The next Fig.~\ref{fig5_ampl} presents a dependence of the maximal SR coefficient $\delta R_\text{fix}$ on pump intensity for four atomic densities. The increase of the pump intensity results in reduction of $\delta R_\text{fix}$ and the power of the reflected probe beam. Discussed research extends our previous results~\cite{sautenkov2021optical, bobrov2021dipole, sautenkov2022spectral}, obtained with a single laser beam.

\begin{figure}[ht]
    \includegraphics[width=\linewidth]{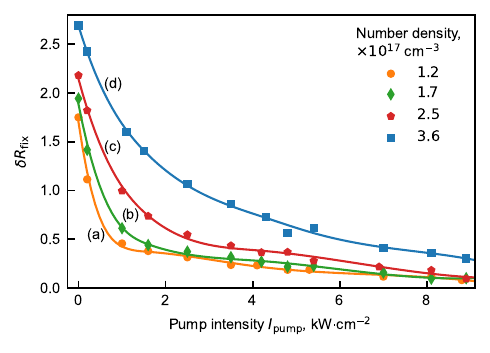}
    \caption{Normalized SR coefficient for selected probe frequency (dashed line in Fig.~\ref{fig2_SR}) with zero pump laser detuning $\Delta_{\text{pump}}/2\pi=0$. The solid curves are guides to the eyes. (a) The set of measurements (orange circles and the curve) has been performed at the density of $1.2\times10^{17}$~cm$^{-3}$; (b) the green diamonds and the curve are for $1.7\times10^{17}$~cm$^{-3}$; (c) the red pentagons and the curve are for $2.5\times10^{17}$~cm$^{-3}$; (d) the blue squares and the curve are for $3.6\times10^{17}$~cm$^{-3}$.}
    \label{fig5_ampl}
\end{figure}

The use of a pulsed pump beam allows to develop a prototype of an all optical switch or modulator. We would like to note that in the current optical scheme involving nearly normal incident angles for the pump and probe optical beams, there is an offset due to non-resonant reflection from the interface window/atomic vapor. The offset reduces the contrast of the signal. The contrast of the modulation can be enhanced by using the probe beam at a large incident angle~\cite{akulshin1989selective,sautenkov2011variable}.

An extension of the nonlinear spectroscopy technique to thin cells~\cite{keaveney2012maximal, peyrot2018collective, sargsyan2023competing} will open new opportunities in research and in application. In recently published paper~\cite{christaller2022transient}, transient density-induced dipolar interactions were investigated in a micrometer-sized cell with rubidium vapor. Atomic desorption by off-resonant laser pulse was used to produce high atomic densities. Transient evolution of dipole broadening and line shift were measured. In~\cite{christaller2022transient} it was assumed that fast switching of atomic density could be used for the development of future quantum devices.  
We suggest to apply for controlling optical properties of the vapor thin cells the pulse optical saturation instead of the off-resonance light-induced atomic desorption. The optical saturation technique can be more flexible in comparison with laser desorption. An important difference between the techniques exists. The desorption laser pulse enhances the dielectric constant and dipole--dipole interactions. The saturation pump pulse diminishes the dielectric constant and dipole--dipole interactions. It is a challenge to conduct the suggested research for thin vapor cells.

\section{Conclusions}

Nonlinear selective reflection from the interface YAG window--high density rubidium is investigated using pump-probe scheme. Growth of the pump beam intensity causes reduction of the magnitude and the width of the selective reflection resonances. By changing the pump intensity, it is possible to control the spectral width and reflectivity. We assume that nonlinear selective reflection can be used in order to develop a fast all-optical modulator.

\section*{CRediT authorship contribution statement}

\textbf{Vladimir Sautenkov}: Conceptualization, Methodology, Validation, Writing -- original draft, Funding acquisition. \textbf{Sergey Saakyan}: Investigation, Software, Validation, Formal analysis, Writing -- review \& editing. \textbf{Andrei Bobrov}: Formal analysis, Methodology, Writing -- review \& editing. \textbf{Leonid Khalutornykh}: Investigation, Writing -- review \& editing. \textbf{Boris B. Zelener}: Supervision, Resources, Writing -- review \& editing, Funding acquisition.

\section*{Declaration of Competing Interest}
The authors declare that they have no known competing financial interests or personal relationships that could have appeared to influence the work reported in this paper.

\section*{Data availability}
Data will be made available on request.

\section*{Acknowledgements}
The research has been supported by the Russian Science Foundation Grant No.\,23-22-00200. The development of the experimental setup for high-density vapor investigations was supported by the Ministry of Science and Higher Education of the Russian Federation (State Assignment No.\,075-00270-24-00).

\bibliographystyle{elsarticle-num}
\bibliography{refs}

\end{document}